%% file: main.tex
\documentclass[lettersize,journal]{IEEEtran}

\input{pms}

\begin{document}

\title{When Majority-Vote Fails:  
Sequence Reconstruction under Fixed-Weight Substitutions}

\author{%
  Vivian~Papadopoulou\,\orcidlink{0009-0005-8593-974X},\,%
  Yonatan~Yehezkeally\,\orcidlink{0000-0003-1652-9761}\,%
		,~\IEEEmembership{Member,~IEEE,}\\
  V.~Arvind~Rameshwar\,\orcidlink{0000-0002-5284-215X}\,%
		,~\IEEEmembership{Member,~IEEE,}
  Antonia~Wachter-Zeh\,\orcidlink{0000-0002-5174-1947}\,%
		,~\IEEEmembership{Senior~Member,~IEEE}%
  \thanks{%
  Vivian~Papadopoulou and Antonia~Wachter-Zeh are with the 
  Technical University of Munich, School of Computation, Information and Technology, Munich, Germany 
  (e-mails: \texttt{\{vivian.papadopoulou,antonia.wachter-zeh\}@tum.de}).
  Vivian~Papadopoulou is also with 
  Imperial College London, Department of Electrical and Electronic Engineering, London, United Kingdom.
  Yonatan~Yehezkeally is with the 
  ICOS Research Group, School of Computing, Newcastle University (UK), Newcastle upon Tyne NE1 7RU, United Kingdom 
  (e-mail: \texttt{yonatan.yehezkeally@ncl.ac.uk}).
  V.~Arvind~Rameshwar is with the Department of Electrical Engineering,
  Indian Institute of Technology Madras,  Chennai, India
  (e-mail: \texttt{arvind@ee.iitm.ac.in}).}%
}

\maketitle

\begin{abstract}

\input{chapters/0_abstract}
\end{abstract}


\vspace{0.5cm}

\section{Introduction}
\input{chapters/1_intro}

\section{Preliminaries \& System Setup}\label{sec:system_setup}
\input{chapters/2_prelims}

\section{Majority-Vote Decoding: Guarantees and Limitations}\label{sec:majority}
\input{chapters/3_majority}

\section{Weighted Majority-Vote Decoders}\label{sec:wmvd}
\input{chapters/4_wmvd}

\section{Conclusion}
\input{chapters/7_conc}

\section*{Acknowledgments}
\noindent The authors would like to thank Christoph Hofmeister for meaningful discussions and comments.

\section*{AI Statement}

Authors used Claude (Anthropic) to assist in finding efficiencies for the code for the exhaustive enumerations of Table~\ref{tab:backtracking_performance} and the Monte Carlo simulation performed to generate Fig.~\ref{fig:early_term}.

\printbibliography

\end{document}

%% file: pms.tex
\usepackage{amsmath,amsfonts,amssymb, amsthm}
\usepackage{dsfont}
\usepackage{mathtools}
\usepackage{algorithmic}
\usepackage[dvipsnames]{xcolor}
\usepackage{algorithm}
\usepackage{array}
\usepackage[caption=false,font=normalsize,labelfont=sf,textfont=sf]{subfig}
\usepackage{textcomp}
\usepackage{stfloats}
\usepackage{url}
\usepackage{verbatim}
\usepackage{graphicx}
\usepackage{booktabs}
\usepackage{tabularx} 
\usepackage{algorithm}
\usepackage{algorithmic}
\usepackage{xcolor}
\usepackage{makecell}
\usepackage{colortbl} 
\usepackage{multirow}
\usepackage{booktabs}
\usepackage{orcidlink} 

\usepackage[backend=biber,
            style=ieee,
            citestyle=numeric-comp,
            isbn=false,
            hyperref=false]{biblatex} 
\AtEveryBibitem{%
  \ifentrytype{online}{}{%
    \ifentrytype{unpublished}{}{%
      \clearfield{urldate}%
      \clearfield{urlyear}%
      \clearfield{urlmonth}%
      \clearfield{urlday}%
    }%
  }%
}

\newtheorem{theorem}{Theorem}

\newtheorem{lemma}{Lemma}
\newtheorem{remark}{Remark}
\newtheorem{example}{Example}
\newtheorem{corollary}[theorem]{Corollary}
\newtheorem{definition}{Definition}

\newtheorem{conjecture}{Conjecture}

\DeclareMathOperator{\wt}{wt}
\newcommand{\NLev}{N_{\mathrm{Lev}}}
\newcommand{\Nmaj}{N_{\mathrm{maj}}}
\newcommand{\Nwc}{N_{\mathrm{wc}}}
\newcommand{\Nmin}{N_{\mathrm{min}}}
\newcommand{\Nwmaj}{N_{\mathrm{wmaj}}}
\newcommand{\wtH}{\wt_{\mathrm{H}}}
\newcommand{\dH}{d_{\mathrm{H}}}
\newcommand{\Ftwo}{\mathbb{F}_2}

%% file: chapters/0_abstract.tex
We consider the problem of binary sequence reconstruction where a sequence is transmitted multiple times over a channel, and each undergoes substitution errors by an arbitrary error vector of Hamming weight exactly~$t$. 
We determine $\Nwc$, the minimum number of \emph{distinct} channel output observations \emph{guaranteeing} unique reconstruction by an optimal decoder, and show that the bit-wise majority-vote decoder, which was proved by Levenshtein (2001) to be optimal for channels corrupting \emph{up to}~$t$ bits, is in fact suboptimal here. 
Despite this suboptimality, we prove a condition that admits a practical early-termination rule, necessary for every decoder at any number of outputs $N$, and sufficient for bit-wise majority-vote even below~$\Nwc$. We evaluate the number of outputs at which this condition is triggered, numerically, and show that it frequently occurs below $\Nwc$ when the block-length~$k>2t+1$. 
Finally, we propose an alternative algorithm aimed at guaranteeing unique reconstruction with fewer observations than bit-wise majority-vote, and demonstrate through exhaustive enumeration that it successfully decodes channel output-sets where bit-wise majority-vote fails.

%% file: chapters/1_intro.tex
\IEEEPARstart{T}{he} problem of sequence reconstruction, first formalized by Levenshtein in \cite{Levenshtein2001,Levenshtein2001a}, addresses the recovery of an unknown sequence from multiple noisy channel output observations. The setup models settings in which redundancy introduced at the source might be insufficient for error-correction. 
Among other error types, Levenshtein determined the minimum number of \emph{distinct} channel output observations required by \emph{any} decoder to \emph{guarantee} unique reconstruction when each observation is corrupted by up-to-$t$ substitution errors. 
Levenshtein also showed that the bit-wise majority-vote decoder is optimal, in that it successfully reconstructs given that many distinct observations. 
Since then, Levenshtein's work was further extended in the case of deletion and insertion errors \cite{Gabrys2018,Junnila2025,Lan2026}, mixed error types \cite{Cai2022,Junnila2025}, coded settings \cite{Abu-Sini2021,Cai2022} where the transmitted sequence belongs to a code, and list decoding \cite{YaaBru19,YehSch21,Junnila2024,Abu-Sini2024} where the decoder may output multiple candidates.

The works above establish worst-case guarantees under a combinatorial channel model, where the decoder must succeed given any set of sufficiently many channel outputs. A complementary line of work studies probabilistic channel models, e.g., \cite{De2017,Holden2018, Rameshwar2025}, where the goal is to determine how many channel output observations guarantee reconstruction with high probability. Recently, alternative perspectives have also emerged within Levenshtein's combinatorial framework. In \cite{Wang2024, Wang2026}, sufficient conditions for unique reconstruction that depend not only on the number of observations, but also on the pairwise distances among them are derived, enabling reconstruction with fewer observations than the worst-case threshold.

A variant of Levenshtein's channel, in which each observation is corrupted by \emph{exactly}~$t$ substitution errors, rather than \emph{at most}~$t$, was introduced in \cite{Papadopoulou2024}, where the average-case performance of the bit-wise majority-vote decoder was studied and bounds on the expected number of distinct noisy channel output observations needed for reconstruction under uniformly random constant-weight error patterns were derived. This model serves as a structured restriction of the classical up-to-$t$ model that renders the average-case analysis tractable. 
Beyond its analytical tractability, the exact-$t$ model is also motivated by settings where the error weight per observation is determined by a system-level constraint rather than a probabilistic mechanism, e.g., in adversarial scenarios where a jammer with a fixed corruption budget utilizes all available resources per transmission.

In the present work, we study the worst-case behaviour of this exact-$t$ substitution channel and show that, perhaps surprisingly, the additional structure imposed by the fixed error weight renders the bit-wise majority-vote decoder suboptimal, thereby calling for the design of alternative reconstruction algorithms. 

The remainder of this paper is organized as follows. 
Section~\ref{sec:system_setup} introduces necessary notation, relevant literature, and sets up the exact-$t$ substitution channel. Section~\ref{sec:majority} presents our results on bit-wise majority-vote suboptimality, by identifying the reconstruction-guarantee threshold, establishing a frequently encountered early-termination rule, and evaluating with representative parameters the performance gap between an optimal decoder and the bit-wise majority-vote decoder, which validates our claims. 
Section~\ref{sec:wmvd} introduces a proposed weighted majority-vote decoder, studies its behaviour, and presents numerical evaluations supporting it as candidate decoder in regimes where bit-wise majority fails to reconstruct the transmitted sequence.

%% file: chapters/2_prelims.tex
\subsection{Notation}
We use lowercase math font for scalars (e.g., $a$) and lowercase bold math font for vectors (e.g., $\mathbf{a}$). The $j$-th entry of a vector $\mathbf{a}$ is written $\mathbf{a}[j]$. 
Uppercase letters (e.g., $Y$) denote sets. As an exception, and to align with existing literature, we use $N$ (including subscripts) to denote specific scalar integers throughout the paper.

We write $[n]$ for the set of positive integers $\{1,\ldots,n\}$, and denote the binary field by $\Ftwo$. The Hamming weight of a vector $\mathbf{a} \in \Ftwo^k$ is denoted $\wtH(\mathbf{a})$, and the Hamming distance between two vectors $\mathbf{a}, \mathbf{b} \in \Ftwo^k$ is denoted $\dH(\mathbf{a},\mathbf{b})$. The Hamming sphere of radius $t$ centred at $\mathbf{u} \in \Ftwo^k$ is defined as $S_t(\mathbf{u}) = \left\{ \mathbf{y} \in \Ftwo^k : \dH(\mathbf{u}, \mathbf{y}) = t \right\}$, with cardinality $|S_t(\mathbf{u})|=\binom{k}{t}$. We denote $\Ftwo^k$ addition, i.e., the bit-wise exclusive-OR (XOR) operation, by $\oplus$. We also write $\mathds{1}\{\cdot\}$ for the indicator function, which equals $1$ when its argument is true and $0$ otherwise.
 
\subsection{System Setup and Channel Model}

Consider the transmission of a binary sequence $\mathbf{u} \in \Ftwo^k$ over the following channel.
\begin{definition}\label{def:channel}
Given an integer $t\geq1$ and an arbitrary input vector $\mathbf{u} \in \Ftwo^k$, the binary \emph{exact-$t$-substitution channel}, denoted by $W_t^{(N)}$, outputs $N$ distinct binary sequences \{$\mathbf{y}_i \in S_t(\mathbf{u}) : i \in [N]$\}.
\end{definition}
\noindent 
The distinctness of the channel outputs implies $N \leq \binom{k}{t}$. 
Throughout this manuscript, we restrict our attention to $t<\frac{k}{2}$; the case of $t>\frac{k}{2}$ reduces to the former by considering the bitwise complement of each $\mathbf{y}_i$. In the case $k=2t$, it is unfortunately insufficient even to acquire $N=|S_t(\mathbf{u})|=\binom{k}{t}$ unique outputs, since $S_t(\mathbf{u}) = S_t(\bar{\mathbf{u}})$, where $\bar{\mathbf{u}}$ is the bit-wise complement of $\mathbf{u}$. Thus, unique non-coded reconstruction is an impossible proposition, and we do not treat it in this manuscript.

When expedient, and if it can be done without loss of generality, 
we may assume that the transmitted vector $\mathbf{u}=\mathbf{0}$, the all-zeros vector.
\begin{definition}\label{def:seqmatrix}
Let 
\begin{align*}
    Y_N(k,t) \triangleq \lbrace Y\subseteq S_t(\mathbf{0})\subseteq \Ftwo^k : |Y| = N\rbrace 
\end{align*}
be the collection of all $N$-subsets of the $t$-sphere around $\mathbf{0}\in \Ftwo^k$. When convenient we may think of such sets $Y$ as matrices (up to row-permutations), and refer to $\lbrace \mathbf{y}_i[j] : i\in [N]\rbrace$ as the $j$th column of~$Y$. Note that in such a matrix all rows are distinct and have Hamming weight exactly $t$. 
We follow \cite[Def.~1]{Papadopoulou2024} in referring to such a matrix as a \emph{sequence matrix}. 
For column $j \in [k]$, we denote the \emph{column sum} 
\begin{align*}
    c_j\triangleq |\lbrace i\in[N] : \mathbf{y}_i[j]\neq \mathbf{u}[j]\rbrace|, 
\end{align*}
i.e., the number of channel outputs in which position $j$ is in error. Note that $\sum_{j=1}^{k} c_j = tN$.
\end{definition}

\subsection{Majority-Vote Decoder and Known Results}
A decoder for the exact-$t$-substitution channel is a mapping from the
$N$ distinct channel outputs to an estimate $\hat{\mathbf{u}} \in \Ftwo^k$ of the transmitted sequence. We say that a decoder \emph{successfully reconstructs} $\mathbf{u}$ if
$\hat{\mathbf{u}} = \mathbf{u}$.

The bit-wise majority-vote decoder estimates each position of $\mathbf{u}$ independently by selecting the value that appears most frequently at that position across $N$ observations. Position~$j$ is decoded correctly when the column sum satisfies $c_j < \frac{N}{2}$, since the majority of observations then agree on the correct value. Formally, the decoder outputs:
\begin{align}
\label{eq:majrule}
\hat{\mathbf{u}}[j] \triangleq \operatorname*{argmax}_{x\in \Ftwo} \big|\lbrace \mathbf{y}_i[j] = x : i\in N\rbrace\big|,
\end{align}
or $\circledast$ in case of ties, denoting an erasure (decoding failure). The erasure case can only occur when $N$ is even; to avoid this, we assume throughout that $N \geq 3$ is odd.

In Levenshtein's classical setting, where each observation may contain \textit{at most}~$t$ substitution errors (i.e., $0 \leq \dH(\mathbf{y}_i,\mathbf{u}) \leq t$), the minimum number of distinct channel outputs required to guarantee reconstruction over a
$q$-ary alphabet was derived in~\cite{Levenshtein2001}. For the binary case ($q=2$), this expression simplifies to 
\begin{align*}
    \NLev = 2\sum_{i=0}^{t-1}\binom{k-1}{i} + 1.
\end{align*}
Levenshtein showed that the majority-vote decoder achieves this bound, meaning that it is worst-case optimal for this up-to-$t$-substitution channel: no decoder, regardless of complexity, can \textit{guarantee} reconstruction with fewer than $\NLev$ observations.

For the exact-$t$-substitution channel, the worst-case number of observations required by the majority-vote decoder was established in~\cite{Papadopoulou2024}.

\begin{theorem}\cite[Theorem~1]{Papadopoulou2024}\label{th:Nmaj}
The minimum number of distinct outputs of the exact-$t$-substitution channel required to guarantee reconstruction of a binary sequence $\mathbf{u} \in \Ftwo^k$ by the bit-wise majority-vote decoder is
\begin{align*}
    \Nmaj = 2\binom{k-1}{t-1} + 1.
\end{align*}
\end{theorem}
We note that this result was stated as $\Nwc$ (`worst case') in~\cite{Papadopoulou2024}; we use $\Nmaj$ here to emphasize that it is the worst case for the majority-vote decoder specifically, and reserve the notation $\Nwc$ to refer to \emph{any} possible decoder. Since $\binom{k-1}{t-1} \leq \sum_{i=0}^{t-1}\binom{k-1}{i}$, we have $\Nmaj \leq \NLev$, which confirms the fact that the exact-$t$-substitution channel does not increase the worst-case requirement for majority-vote decoding relative to the classical up-to-$t$ setting. Equality holds if and only if $t=1$, in which case $\Nmaj=\NLev=3$. For $t\geq2$, the inequality is strict, indicating as expected that the exact-$t$ constraint already reduces the worst-case requirement for majority-vote decoding compared to the classical setting.

Unlike in the classical setting, however, the majority-vote decoder is surprisingly not optimal for the exact-$t$-substitution channel. In Section~\ref{sec:majority}, we establish the minimum number of observations $\Nwc$ required by \textit{any} decoder to guarantee reconstruction, and show that $\Nwc < \Nmaj$ for all $t \geq 2$. Table~\ref{tab:reconstruction_bounds}) summarises these thresholds for the exact-$t$-substitutions channel.

\begin{table*}[t]
\centering
\caption{Summary of Fundamental Limits/ Thresholds}
\label{tab:reconstruction_bounds}

\renewcommand{\arraystretch}{1.2} 
\begin{tabularx}{\textwidth}{p{1cm} p{2cm} p{2cm} X}
\toprule
\textbf{Notation} & \textbf{Bound} & \textbf{Source} & \textbf{Description and Significance} \\ 
\midrule

$\Nwc$ & $2 \binom{k-2}{t-1} + 1$ & Th.~\ref{th:universalN} & 
\textbf{Reconstruction Threshold (Exact-$t$)}: Minimum number of distinct views guaranteeing unique reconstruction. \\ 
\addlinespace[2.5ex]

$\Nwmaj$ & $\Big\lfloor\frac{\binom{k-1}{t-1}}{1-\frac{t}{k}}\Big\rfloor+1$ & Eq.~\eqref{eq:Nwmaj} & 
\textbf{Weighted-Majority-Vote Threshold (Exact-$t$)}: Minimum number of distinct views guaranteeing successful reconstruction in the first step by the decoder of Alg.~\ref{alg:weighted}. \\
\addlinespace[2.5ex]

$\Nmaj$ & $2 \binom{k-1}{t-1} + 1$ & \cite[Th.~1]{Papadopoulou2024} & 
\textbf{Majority-Vote Threshold (Exact-$t$)}: Minimum number of distinct views guaranteeing unique reconstruction by the standard bit-wise majority-vote decoder.  \\ 

\bottomrule
\end{tabularx}
\end{table*}

%% file: chapters/3_majority.tex
In this section, we determine the minimum number~$\Nwc$ of distinct channel outputs for which unique reconstruction is possible by \textit{any} decoder, and show that this quantity is strictly smaller than the majority-vote guarantee~$\Nmaj$. We also derive an error detection property that provides a sufficient condition for verifying correct reconstruction without knowledge of the transmitted sequence. 

\subsection{Worst-case reconstruction guarantee threshold}
Given a set of distinct channel outputs $Y=\{\mathbf{y}_i : i \in [N]\}$, the transmitted sequence may be any element of 
\begin{align*}
U \triangleq \bigcap_{\mathbf{y} \in Y} S_t(\mathbf{y}).
\end{align*}
i.e., the set of all binary sequences that are at Hamming distance exactly $t$ from every observation. Error-free reconstruction is \emph{guaranteed} only when $U$ is a singleton. For any pair of distinct candidate sequences $\mathbf{u}_1\neq \mathbf{u}_2$, we denote the set of channel outputs that are consistent with both candidates, i.e., the \emph{ambiguity set}, as:
\begin{multline*}
    S_t(\mathbf{u}_1) \cap S_t(\mathbf{u}_2) \\
    = \{\mathbf{y} \in \Ftwo^k : \dH(\mathbf{y}, \mathbf{u}_1) = \dH(\mathbf{y}, \mathbf{u}_2) = t\}.
\end{multline*}
Note that $U$ is an intersection of spheres around all observations, while the ambiguity set is an intersection of spheres around two candidates. 
If all $N$ observations happen to fall in the ambiguity set, then the decoder cannot distinguish $\mathbf{u}_1$ from $\mathbf{u}_2$. Therefore, guaranteeing $|U|=1$ requires
\begin{equation*}
    N > \max_{\mathbf{u}_1\neq \mathbf{u}_2 \in \Ftwo^k} \left| S_t(\mathbf{u}_1) \cap S_t(\mathbf{u}_2) \right|,
\end{equation*}
i.e., the number of observations must exceed the size of the largest ambiguity set over all pairs of candidate sequences. Conversely, this condition is also sufficient, as it ensures that no two candidates can both be consistent with all observations. Therefore
\begin{align}\label{eq:Nwc_general}
    \Nwc = 1 + \max_{\mathbf{u}_1\neq \mathbf{u}_2 \in \Ftwo^k} \left| S_t(\mathbf{u}_1) \cap S_t(\mathbf{u}_2) \right|.
\end{align}

In order to find that size, we repeat a well-known lemma, and for completeness prove it below.
\begin{lemma}\label{lem:sphere-intersection}
Let $\mathbf{u}_1, \mathbf{u}_2, \mathbf{y} \in \Ftwo^k$ satisfy $\dH(\mathbf{y}, \mathbf{u}_1) = \dH(\mathbf{y}, \mathbf{u}_2) = t$. Denote $D \triangleq \{i \in [k] : \mathbf{u}_1[i] \neq \mathbf{u}_2[i]\}$. Then,
\begin{align*}
\left| \{i \in D : \mathbf{y}[i] = \mathbf{u}_1[i]\} \right| = \left| \{i \in D : \mathbf{y}[i] = \mathbf{u}_2[i]\} \right| = \frac{1}{2} |D|.
\end{align*}
\end{lemma}

\begin{IEEEproof}
Observe that
\begin{align*}
\{i \not\in D : \mathbf{y}[i] \neq \mathbf{u}_1[i]\} &= \{i \not\in D : \mathbf{y}[i] \neq \mathbf{u}_2[i]\}, \\
\{i \in D : \mathbf{y}[i] \neq \mathbf{u}_2[i]\} &= D \setminus \{i \in D : \mathbf{y}[i] \neq \mathbf{u}_1[i]\}.
\end{align*}
It therefore suffices to prove the left-most equality in the lemma's statement. 
Then,
\begin{align*}
&|\{i \not\in D : \mathbf{y}[i] \neq \mathbf{u}_1[i]\}| + |\{i \in D : \mathbf{y}[i] \neq \mathbf{u}_1[i]\}| \\
&= \dH(\mathbf{y}, \mathbf{u}_1) = t = \dH(\mathbf{y}, \mathbf{u}_2) \\
&= |\{i \not\in D : \mathbf{y}[i] \neq \mathbf{u}_2[i]\}| + |\{i \in D : \mathbf{y}[i] \neq \mathbf{u}_2[i]\}| \\
&= |\{i \not\in D : \mathbf{y}[i] \neq \mathbf{u}_1[i]\}| + |D| - |\{i \in D : \mathbf{y}[i] \neq \mathbf{u}_1[i]\}|.
\end{align*}
It follows that $|D| = 2|\{i \in D : \mathbf{y}[i] \neq \mathbf{u}_1[i]\}|$, which concludes the proof.
\end{IEEEproof}

\begin{theorem}\label{th:universalN}
The minimum number $\Nwc$ of distinct outputs of the exact-$t$-substitution channel required to guarantee reconstruction of a transmitted binary sequence $\mathbf{u} \in \Ftwo^k$ when $t<\frac{k}{2}$ is given by:
\begin{align*}
    \Nwc = 2\binom{k-2}{t-1} + 1.
\end{align*}
\end{theorem}
\begin{IEEEproof}
Fix two distinct $\mathbf{u}_1, \mathbf{u}_2\in \Ftwo^k$. Denote $d \triangleq \frac{1}{2} \dH(\mathbf{u}_1, \mathbf{u}_2) = \frac{1}{2} |D|>0$. By Lemma~\ref{lem:sphere-intersection}, any $\mathbf{y} \in S_t(\mathbf{u}_1) \cap S_t(\mathbf{u}_2)$ must disagree with each of $\mathbf{u}_1, \mathbf{u}_2$ on exactly $d$ coordinates in $D$, and with both of them on exactly $t - d$ coordinates in $[k] \setminus D$ (by the triangle inequality, no such $\mathbf{y}$ can exist for $d>t$). Conversely, any choice of $d$ coordinates in $D$ and $t - d$ coordinates in $[k] \setminus D$ yields a unique $\mathbf{y}$ satisfying $\dH(\mathbf{y}, \mathbf{u}_1) = \dH(\mathbf{y}, \mathbf{u}_2) = t$. Substituting into~\eqref{eq:Nwc_general}, it follows that
\begin{align*}
    \Nwc = \max_{d\in [t]} \binom{2d}{d} \binom{k - 2d}{t - d} + 1.
\end{align*}

In order to find the maximum, we denote $f_{k,t}(d) = \binom{2d}{d} \binom{k - 2d}{t - d}$ for $d\in[t]$. We will show that $f_{k,t}(d)\leq f_{k,t}(1)$; equivalently, that $q_{k,t}(d)\triangleq \frac{f_{k,t}(d)}{f_{k,t}(1)}\leq 1$.

Observe that 
\begin{align*}
    \frac{q_{k+1,t}(d)}{q_{k,t}(d)} &= \frac{(k+1-2d) (k-t)}{(k+1-d-t) (k-1)},
\end{align*}
thus $q_{k+1,t}(d)\leq q_{k,t}(d)$ if and only if 
\begin{align*}
    (k+1-2d) (k-t)\leq (k+1-d-t) (k-1).
\end{align*}
Rearranging gives $(k+1-2t) (d-1)\geq 0$, which holds for all $t < \frac{k}{2}$ and $d\in [t]$. Therefore, it suffices to verify $q_{2t+1,t}(d)\leq 1$ (since $t<\frac{k}{2}$ implies $k\geq 2t+1$).

For that purpose, denote 
\begin{align*}
    r_{k,t}(d) 
    \triangleq \frac{f_{k,t}(d+1)}{f_{k,t}(d)} 
    = \frac{2 (2d+1) (t-d) (k-d-t)}{(d+1) (k-2d) (k-2d-1)}, 
\end{align*}
for $d\in[t-1]$. Observe 
\begin{align*}
    r_{2t+1,t}(d) 
    = \frac{(2d+1) (t+1-d)}{(2d+1) (t+1-d) + (t-2d)}, 
\end{align*}
hence, $f_{2t+1,t}(d)$ is decreasing for $d\leq \frac{t}{2}$, and increasing for $d > \frac{t}{2}$. It therefore achieves its maximum at either $d=1$ or $d=t$, and we note $f_{2t+1,t}(1) = 2\binom{2t-1}{t-1} = \binom{2t}{t} = f_{2t+1,t}(t)$. Thus, $q_{k,t}(d)\leq q_{2t+1,t}(d)\leq 1$, and $f_{k,t}(d)$ is maximized for $d=1$.
\end{IEEEproof}

By Pascal's identity we have $\Nmaj - \Nwc = 2\binom{k-2}{t-2} > 0$ for all $2\leq t < \frac{k}{2}$. This confirms that the bit-wise majority-vote decoder is suboptimal for the exact-$t$-substitution channel: the transmitted sequence can uniquely be determined from any set of channel outputs of size $\Nwc \leq N < \Nmaj$, but the majority-vote decoder may fail on some such choices (indeed, we will see later in Table~\ref{tab:backtracking_performance} that this occurs in practice). For $t = 1$, both expressions yield $\Nwc = \Nmaj = 3$, consistent with~\cite[Proposition~2]{Papadopoulou2024}. Thus, the gap between $\Nwc$ and $\Nmaj$ exists only for $t \geq 2$.

In order to quantify the gain of decoding with $\Nwc$, instead of $\Nmaj$, distinct channel output views, we note that not only does the gap~$\Nmaj - \Nwc$ grow with~$t$, but also 
\begin{align*}
    \frac{\Nwc}{\Nmaj - \Nwc} = 1 + \frac{k-2t+1}{t-1} + \frac{1}{2\binom{k-2}{t-2}}
\end{align*}
decreases, thus the gap increases \emph{as a fraction} of $\Nwc$ (for large~$k$, approaching $\Nwc$ as~$t$ approaches $\frac{k}{2}$). Therefore, decoders that can guarantee reconstruction with $\Nwc$ distinct observations, or even a number $\Nwc\leq N < \Nmaj$, become highly desirable. 
Table~\ref{tab:Gap} demonstrates this rule for a specific private case, for fixed $k=20$ and all $2\leq t < \frac{k}{2}$.

\begin{table}[H]
\centering
\caption{Example instances of the $\Nmaj - \Nwc$ gap for fixed $k=20$ and various $t$ values up to the border case $t=\big\lfloor\frac{k-1}{2}\big\rfloor$}\label{tab:Gap}

\begin{tabular}{c c c r r r}
\toprule
$t$ & $\Nwc$ & $\Nmaj$ & Gap  & Relative Gap 
\\\midrule
2 & 37 & 39 & 2 & 5.4\% 
\\3 & 307& 343 & 36 & 11.7\% 
\\4 & 1{,}633 & 1{,}939 & 306 & 18.7\%  
\\5 & 6{,}121 & 7{,}753 & 1{,}632 & 26.7\%  
\\6 & 17{,}137 & 23{,}257 & 6{,}120 & 35.7\%  
\\7 & 37{,}129 & 54{,}265 & 17{,}136 & 46.2\%  
\\8 & 63{,}649 & 100{,}777 & 37{,}128 & 58.3\% 
\\9 & 87{,}517 & 151{,}165 & 63{,}648 & 72.7\%  
\\
\bottomrule
\end{tabular}
\end{table}

\subsection{Error detection and termination of bit-wise majority-vote decoding}
When $N < \Nmaj$, the bit-wise majority-vote decoder may still, \textit{sometimes}, produce the correct output $\hat{\mathbf{u}} = \mathbf{u}$. In such cases, it is desirable for the decoder to detect this success and terminate early, rather than requiring $\Nmaj$ observations.

By the definition of the exact-$t$-substitution channel, $\dH(\mathbf{y}_i, \mathbf{u}) = t$ for all $i \in [N]$. Conversely, if $\dH(\mathbf{y}_i, \hat{\mathbf{u}}) \neq t$ for at least one $i$, then $\hat{\mathbf{u}} \neq \mathbf{u}$, indicating a detectable error. 
Theorem~\ref{th:undetectable} establishes that this condition is not only necessary but also sufficient for unique decoding by majority-vote: no ambiguity can be missed.

\begin{theorem}\label{th:undetectable}

For any pair $(k,t)$ such that $0 < t < \frac{k}{2}$, suppose that an odd number $1<N<\Nmaj$ of distinct channel output observations are obtained, and let $\hat{\mathbf{u}}$ be their point-wise majority. If $\dH(\mathbf{y}_i, \hat{\mathbf{u}}) = t$ for all $i \in [N]$, then $\hat{\mathbf{u}} = \mathbf{u}$.
\end{theorem}

\begin{IEEEproof}
Recall by assumption that for all $i \in [N]$ it holds that $\dH(\mathbf{y}_i, \mathbf{u}) = t$, and denote $\mathbf{z}_i\triangleq \mathbf{y}_i\oplus\mathbf{u}$ (so that $\wt_H(\mathbf{z}_i) = t$). Trivially, $\{\mathbf{z}_i\}$ are unknown to the decoder. 
For any binary `column' vector $\mathbf{x} \in \Ftwo^N$, let 
\begin{align*}
    k_\mathbf{x} = \left|\left\lbrace j\in [k] : \forall i\in[N]\colon \mathbf{z}_i[j] = \mathbf{x}[i] \right\rbrace\right|.
\end{align*}
Then,
\begin{align}
\sum_\mathbf{x} k_\mathbf{x} &= k, \label{eq:A}
\end{align}
and for all $i \in [N]$,
\begin{align}
\sum_{\substack{\mathbf{x} \\ \mathbf{x}[i]=1}} k_\mathbf{x} &= t. \label{eq:Bi}
\end{align}

By assumption, for all $i \in [N]$,
\begin{align}
\sum_{\substack{\mathbf{x}:\\ \mathbf{x}[i]=1 \\ \wt_H(\mathbf{x})<\frac{N}{2}}} k_\mathbf{x} +
\sum_{\substack{\mathbf{x}:\\ \mathbf{x}[i]=0 \\ \wt_H(\mathbf{x})>\frac{N}{2}}} k_\mathbf{x}
= \sum_{\substack{\mathbf{x}:\\ \mathbf{x}[i]\neq \operatorname{maj}(\mathbf{x})}} k_\mathbf{x} = t, \label{eq:Ci}
\end{align}
where $\operatorname{maj}(\mathbf{x})$ is the majority bit of the column-vector $\mathbf{x}$.

Subtracting \eqref{eq:Ci} from \eqref{eq:Bi} and dividing by 2, we obtain for each $i \in [N]$:
\begin{align}
&\frac{1}{2} \sum_{\substack{\mathbf{x}\\ \wt_H(\mathbf{x})>\frac{N}{2}}} (2\mathbf{x}[i]-1) k_\mathbf{x} \nonumber\\
&= \frac{1}{2}\Bigg(\sum_{\substack{\mathbf{x}:\\ \mathbf{x}[i]=1 \\ \wt_H(\mathbf{x})>\frac{N}{2}}} k_\mathbf{x} -
\sum_{\substack{\mathbf{x}:\\ \mathbf{x}[i]=0 \\ \wt_H(\mathbf{x})>\frac{N}{2}}} k_\mathbf{x}\Bigg)
= 0. \label{eq:Li}
\end{align}

For each $j \in [N]$, summing \eqref{eq:Li} over all $i \neq j$ yields:

\begin{align}
0 &= \sum_{i\neq j}\frac{1}{2} \sum_{\substack{\mathbf{x}:\\ \wt_H(\mathbf{x})>\frac{N}{2}}} (2\mathbf{x}[i]-1) k_\mathbf{x} \nonumber\\
&= \sum_{\substack{\mathbf{x}:\\ \wt_H(\mathbf{x})>\frac{N}{2}}} k_\mathbf{x} \sum_{i\neq j}(\mathbf{x}[i] - \tfrac{1}{2}) \nonumber\\
&= \sum_{\substack{\mathbf{x}:\\ \wt_H(\mathbf{x})>\frac{N}{2}}} k_\mathbf{x} \big(\wt_H(\mathbf{x}) - \mathbf{x}[j] - \frac{N-1}{2}\big). \label{eq:Ej}
\end{align}

For any $\mathbf{x}$
with $\wt_H(\mathbf{x})>\frac{N}{2}$, we have $\wt_H(\mathbf{x})-\mathbf{x}[j]-\frac{N-1}{2}\geq \frac{N+1}{2}-1-\frac{N-1}{2}=0$. Since $k_\mathbf{x} \geq 0$ by definition, each summand in $\eqref{eq:Ej}$ is non-negative. A sum of non-negative terms equaling zero implies each term is zero, so $k_\mathbf{x}(\wt_H(\mathbf{x}) - \mathbf{x}[j] - \frac{N-1}{2}) = 0$ for all $j \in [N]$ and all $\mathbf{x}$ with $\wt_H(\mathbf{x}) > \frac{N}{2}$.

In particular, for all $\mathbf{x}$ with $\wt_H(\mathbf{x}) > \frac{N+1}{2}$, the factor $\wt_H(\mathbf{x}) - \mathbf{x}[j] - \frac{N-1}{2} > 0$ for every $j$, so $k_\mathbf{x} = 0$. For $\mathbf{x}$ with $\wt_H(\mathbf{x}) = \frac{N+1}{2}$, there exists at least one coordinate $j$ with $\mathbf{x}[j]=0$, giving $\wt_H(\mathbf{x}) - \mathbf{x}[j] - \frac{N-1}{2} = 1 > 0$, so $k_\mathbf{x} = 0$ as well.

Consequently, for all $j\in [k]$ it holds that 
\begin{align*}
    \left|\left\lbrace i\in [N] : \mathbf{z}_i[j]=1 \right\rbrace\right| < \frac{N}{2}.
\end{align*}
Therefore, $\hat{\mathbf{u}} = \mathbf{u}$.
\end{IEEEproof}

\begin{remark}\label{rem:t-codition}
The condition $\dH(\mathbf{y}_i, \hat{\mathbf{u}}) = t$ for all $i \in [N]$ is necessary for $\hat{\mathbf{u}} = \mathbf{u}$ irrespective of the decoder in use, by definition of the channel. Thus, any decoder can use this validity test. When $N\geq \Nwc$, finding any such vector is also sufficient; Theorem~\ref{th:undetectable} is significant because it establishes this as a sufficient condition for the bit-wise majority-vote decoder for \emph{any} odd $N>1$, even when $N<\Nwc$. Note, then, that if $N<\Nwc$ and $\big|\bigcap_{i=1}^N S_t(\mathbf{y}_i)\big|>1$, then Theorem~\ref{th:undetectable} implies that the bit-wise majority $\hat{\mathbf{u}}$ cannot satisfy $\dH(\mathbf{y}_i, \hat{\mathbf{u}}) = t$ for all $i \in [N]$.
\end{remark}

\begin{example}
Let $k=9$, $t=3$. Observe that $\Nwc=43$; nevertheless, let $\mathbf{u}=\mathbf{0}$ be transmitted, and assume a decoder receives 
\begin{align*}
    \mathbf{y}_1 &= (1\; 1\; 1\; 0\; 0\; 0\; 0\; 0\; 0), \\
    \mathbf{y}_2 &= (0\; 0\; 0\; 1\; 1\; 1\; 0\; 0\; 0), \\
    \mathbf{y}_3 &= (0\; 0\; 0\; 0\; 0\; 0\; 1\; 1\; 1).
\end{align*}
Observe, trivially, that $\bigcap_{i\in[3]}S_3(\mathbf{y_i}) = \{\mathbf{u}\}$. Also observe that the bit-wise majority $\hat{\mathbf{u}} = \mathbf{0}$ satisfies $\dH(\hat{\mathbf{u}},\mathbf{y_i})=3=t$ for all $i\in[3]$, thus a decoder can confidently deduce $\hat{\mathbf{u}} = \mathbf{u}$. This, despite observing considerably fewer distinct outputs than $\Nwc$.

In fact, this is trivially the only case of correct decoding by bit-wise majority, up to column ordering (i.e., there are precisely $\frac{1}{3!} \binom{9}{3} \binom{6}{3}$ such output sets, out of a total of $\binom{\binom{9}{3}}{3}$ possibilities).
\end{example}

\begin{figure}[!t]
  \centering
  \includegraphics[width=0.99\columnwidth]{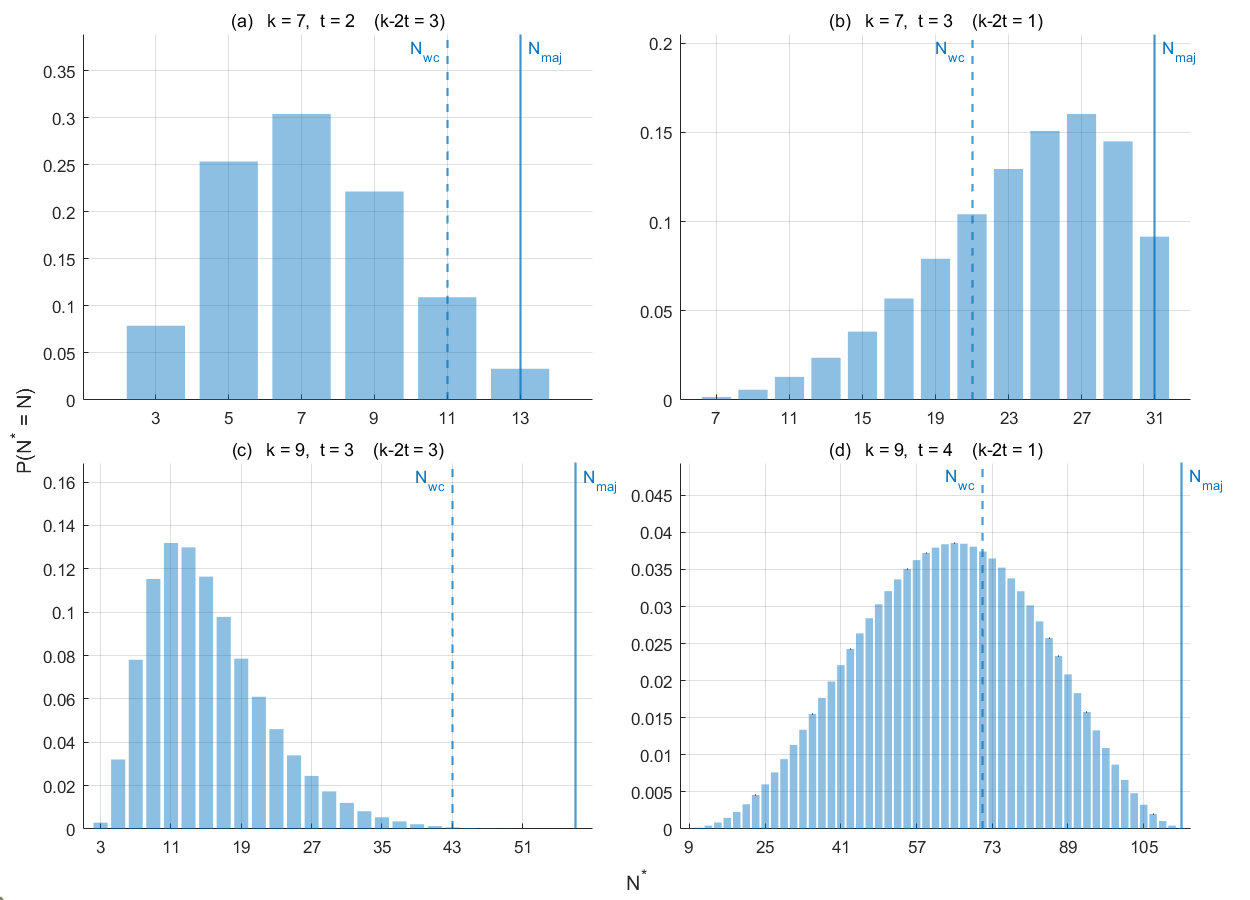}
  \caption{ Monte Carlo estimation of the first instance $N=N^\ast$  where the early-termination rule is triggered during majority-vote decoding. Dashed lines mark $\Nwc$, solid lines mark $\Nmaj$. The first plotted bin in each sub-figure indicates $\Nmin$. Each experiment was run for $10^9$ independent trials. $95\%$ Wilson intervals computed for each bin, with the corresponding error bars being narrower than the bar edges.}
  \label{fig:early_term}
\end{figure}

\begin{remark}\label{rem:Nmin}
Note that the sum of distances $\sum_{i=1}^{N} \dH(\hat{\mathbf{u}}, \mathbf{y}_i) \leq \frac{k(N-1)}{2}$, since by definition of majority-vote each column contributes at most $\frac{N-1}{2}$ disagreements. It follows that for any $\tau \geq 0$, at most $\big\lfloor\frac{k(N-1)}{2\tau}\big\rfloor$ observations can be at Hamming distance $\tau$ or more from $\hat{\mathbf{u}}$.

In particular, since successful reconstruction requires $N$ observations at distance precisely~$t$, we learn that in any instance where bit-wise majority successfully reconstructs $\hat{\mathbf{u}}=\mathbf{u}$, it must hold that $2tN \leq k (N-1)$; this fact was already noted in~\cite[Proposition~1]{Papadopoulou2024}. 
Rearranging gives $N\geq \Nmin \triangleq \big\lceil \frac{k}{k - 2t} \big\rceil$.

For small $t$ relative to $k$, this evaluates to a small number, but as As $t$ approaches $\frac{k}{2}$, the bound grows, indicating that a substantial number of observations is necessary before success is possible. For example, for $k = 10$ and $t = 2$ (where $\Nwc=17$, $\Nmaj=19$), we obtain $\Nmin = 2$, meaning that from the smallest admissible odd $N=3$ onward, success may already be possible. In contrast, for $k=10$ and $t=4$ ($\Nwc=113$, $\Nmaj=169$), we obtain $\Nmin = 5$, so the decoder can be certain that no set of $3$ observations will suffice. This should be taken into consideration when designing when a decoder should in practice utilise the early-termination condition of Theorem~\ref{th:undetectable}.

The three quantities $\Nmin$, $\Nwc$, and $\Nmaj$ together characterize distinct regimes of the reconstruction problem under bit-wise majority-vote decoding: for $N < \Nmin$, no output-set supporting successful decoding can exist; for $\Nmin \leq N < \Nmaj$, success is possible but not guaranteed, and the condition of Theorem~\ref{th:undetectable} serves as an early termination criterion to detect when reconstruction has succeeded; for $N \geq \Nmaj$, success under bit-wise majority-vote is guaranteed. Figure~\ref{fig:early_term} presents Monte Carlo iterations estimating the efficacy of the early-termination condition, including Wilson score $95\%$ confidence intervals. The interesting takeaway is that early termination is not a rarity, even in the most noise-prone settings (when $t=\big\lfloor \frac{k-1}{2}\big\rfloor$).
\end{remark}

\subsection{Counting majority-vote failures at $t=2$, $N=\Nwc$}

In this section we count precisely how frequently bit-wise majority-vote decoding fails at $N=\Nwc$, for $t=2$ (as a first step toward a more general study). We show that the number of output-sets on which this occurs admits a closed form expression in that case.

\begin{definition}\label{def:satcols}
A column $j \in [k]$ of an output-set 
$Y\in Y_N(k,t)$ is \emph{saturated} if $c_j = \binom{k-1}{t-1}$, i.e., if every output which is erroneous at position~$j$ is present in $Y$.
\end{definition}

Since each output $\mathbf{y}_i$ contains $t$ errors, at most $\binom{k-1}{t-1}$ outputs are erroneous at a given position $j$, so $c_j\leq \binom{k-1}{t-1}$. The majority-vote decoder decodes position $j$ incorrectly when $c_j > \frac{N}{2}$, hence for $N=\Nwc=2\binom{k-2}{t-1}+1$ this requires $c_j > \binom{k-2}{t-1}$. Hence for $t=2$ position $j$ is decoded incorrectly if and only if column $j$ is saturated. Saturating a column when $t=2$ therefore forces $k-1$ specific outputs among $N=\Nwc=2k-3$ in total.

\begin{lemma}\label{lem:S2}
For $t=2$, and $N=\Nwc=2k-3$ with $k\geq5$, and any choice of positions $j_1\neq j_2$, there exists a unique sequence matrix $Y\in Y_{\Nwc}(k,2)$ in which both are saturated; we denote that $F_2=1$.
\end{lemma}
\begin{IEEEproof}
We have seen that all $\binom{k-1}{t-1}=k-1$ possible outputs erroneous at either must appear. This includes the unique output erroneous at both, and $2(k-1)$ outputs erroneous at one, but not the other. Since $1+2(k-1)=2k-3=N$, these are precisely all outputs obtained.
\end{IEEEproof}

\begin{corollary}\label{cor:atmost2satcols}
    For $t=2$, and $N=\Nwc=2k-3$, with $k\geq5$, no sequence matrix has three or more saturated columns.
\end{corollary}
\begin{IEEEproof}
By Lemma~\ref{lem:S2}, if positions $j_1\neq j_2$ are saturated, then any other position $j'\notin\{j_1,j_2\}$ is erroneous precisely at a unique output also erroneous at either $j_1,j_2$, i.e., in exactly $2$ outputs. Since $k\geq 5$, no such position is saturated.
\end{IEEEproof}

\begin{lemma}\label{lem:S1}
    For a fixed column $j$, the number of output-sets $Y\in Y_{\Nwc}(k,2)$ in which $j$ is saturated is
    \begin{align*}
        F_1\;=\; \binom{\binom{k-1}{2}}{\,k-2\,}
    \end{align*}
\end{lemma}

\begin{IEEEproof}
  Saturating $j$ forces the $k-1$ outputs erroneous at~
  $j$ to be present. The remaining $N-(k-1) = (2k-3)-(k-1) =k-2$ 
  outputs must be chosen from those error-free at~$j$, of which there are $\binom{k-1}{2}$. The choice is otherwise unconstrained, giving $\binom{\binom{k-1}{2}}{k-2}$ sets.
\end{IEEEproof}

\begin{theorem}\label{th:count}
    For $t=2$, $k\geq 5$ and $N=\Nwc$, the number of sequence matrices on which the bit-wise majority-vote decoder fails is
    \begin{align*}
        F(k)\;=\;k\binom{\binom{k-1}{2}}{\;k-2\;}\;-\;\binom{k}{2}.
    \end{align*}
\end{theorem}

\begin{IEEEproof}
    For $j \in [k]$ let $Y_j$ denote the set of sequence matrices in which column $j$ is saturated. Majority-vote fails exactly on $\bigcup_{j \in [k]} Y_j$. By the inclusion-exclusion principle, summing over all nonempty subsets $T\subseteq [k]$ of columns,
    \begin{align*}
        \Big|\bigcup_{j\in [k]} Y_j\Big| = \sum_{\emptyset\neq T\subseteq[k]}(-1)^{|T|+1}\Big|\bigcap_{j \in T} Y_j\Big|.
    \end{align*}
    By Corollary~\ref{cor:atmost2satcols}, $\bigcap_{j \in T} Y_j=\emptyset$ whenever $|T|\geq3$, so the sum truncates to the first two terms. By Lemma~\ref{lem:S1}, each of the $k$ singleton terms equals $F_1$, and by Lemma~\ref{lem:S2} each of the $\binom{k}{2}$ pair terms equals $F_2=1$. Therefore
    \begin{IEEEeqnarray*}{+rCl+x*}
        F(k)=\binom{k}{1} F_1-\binom{k}{2}F_2 = \binom{k}{1}\binom{\binom{k-1}{2}}{k-2}-\binom{k}{2}
        \\[-\normalbaselineskip] &&&\IEEEQEDhere
    \end{IEEEeqnarray*}
\end{IEEEproof}

\begin{remark}
Theorem~\ref{th:count} reproduces the majority-vote failure counts obtained by exhaustive enumerations in Table~\ref{tab:backtracking_performance}. One may notice that while the absolute numbers are increasing with $k$, they decrease as a fraction of the whole output space $\binom{\binom{t}{k}}{\Nwc}$. This is to be expected in a regime where $t$ is fixed while $k$ increases; future work may quantify precisely the number of point-wise majority vote failures for $t>2$, clarifying the true frequency of failure as, e.g., $t=\Theta(k)$.
\end{remark}

Before concluding this section, we generalise the results of Corollary~\ref{cor:atmost2satcols} as an initial step toward extending the results of this section to $t>2$. For the purposes of this lemma, let $k\geq 2t+1$ and $N=\Nwc$, and assume that $c\geq t$ positions in the bit-wise majority vector $\hat{\mathbf{u}}$ are erroneous. 
For $r\in [t]$, let $N_r\triangleq \binom{c}{r} \binom{k-c}{t-r}$. Take $s\in [t]$ such that 
\begin{align*}
    \sum_{r=s+1}^t N_r < \Nwc\leq \sum_{r=s}^t N_r,
\end{align*}
where the empty sum is understood to be $0$; we know that such~$s$ exists because by the Vandermonde identity 
\begin{align*}
    \sum_{r\in[t]} N_r &= \binom{k}{t}-\binom{k-c}{t} \\
    &= 2\binom{k-2}{t-1} + \left(\binom{k-2}{t}-\binom{k-c}{t}\right) + \binom{k-2}{t-2} \\
    &\geq 2\binom{k-2}{t-1} + 1 = \Nwc.
\end{align*}
Denote $S(k,t,c) \triangleq s\Nwc + \sum_{r=s+1}^t (r-s) N_r$. 

\begin{lemma}\label{lem:max-majority-failures}
We have that
\begin{align*}
    c\leq \frac{S(k,t,c)}{\binom{k-2}{t-1}+1}.
\end{align*}
\end{lemma}
\begin{IEEEproof}
We count instances where any given output is error-free on any of the $c$ positions erroneous in $\hat{\mathbf{u}}$. Note that at most $N_t$ outputs can contain no more than $c-t$ such instances each, because any output cannot be erroneous on more than~$t$ positions. Further, at most $N_{t-1}$ outputs can contain no more than $c-(t-1)$ instances, and so on; that is, 
\begin{align*}
    D &\triangleq \left(\sum_{r=s+1}^t (c-r) N_r\right) + (c-s)\left(\Nwc-\sum_{r=s+1}^t N_r\right)
\end{align*}
bounds from below the count of such instances.

On the other hand, since all $c$ positions are erroneous, each can contribute at most $\frac{\Nwc-1}{2}$ such instances. Therefore, $D\leq c\binom{k-2}{t-1}$. Now, observe 
\begin{align*}
    D 
    &= (c-s)\Nwc - \sum_{r=s+1}^t (r-s) N_r 
    = c\Nwc - S(k,t,c), 
\end{align*}
and rearrangement gives the claim.
\end{IEEEproof}

\begin{remark}
The result of Corollary~\ref{cor:atmost2satcols} can be recovered from Lemma~\ref{lem:max-majority-failures}. Indeed, set $t=2$; if we assume by contradiction that $c=3$ positions can be erroneous, note $N_2 = 3<2k-3=\Nwc$; also $N_1 = 3(k-3)$, hence $N_1+N_2 = 3k-6\geq 2k-3=\Nwc$. Thus $s=1$.

Now, $S(k,2,3) = 3(2k-3) + N_2 = 3(2k-2) > 3(k-1) = 3\big(\binom{k-2}{1}+1\big)$, in contradiction, proving that at most $2$ positions in $\hat{\mathbf{u}}$ may be erroneous.
\end{remark}

%% file: chapters/4_wmvd.tex
\label{sec:wmvd}
The suboptimality of the bit-wise majority-vote decoder established in Section~\ref{sec:majority} motivates the design of a low-complexity decoder that guarantees reconstruction with fewer than $\Nmaj$ observations, perhaps even $\Nwc$. Taking a step towards this goal, in this section we propose a \textit{weighted} majority-vote decoder that exploits the exact-$t$ structure in an attempt to reduce the number of observations required below $\Nmaj$, at the cost of modest complexity increase.

We first introduce the \emph{error-budget} mechanism used in our suggested decoder construction (Section~\ref{subsec:budget}). Our decoder, much like the majority-vote decoder, operates bit-wise; we observe that the order of decoding bits in our decoder is of paramount importance, and we hence aim to decode positions in decreasing order of ``reliability". We analyze the correctness of its first decision, as means of estimating its failure modes (Section~\ref{subsec:firstcol}).  Finally, we explain the order our construction uses, and formally present it (Section~\ref{subsec:order}). The performance of the decoder in comparison to bit-wise majority decoding is then numerically evaluated by exhaustive searches, where it is feasible to do so.

\subsection{Error Budget Tracking and Weights}
\label{subsec:budget}

Bit-wise majority-vote decoding decides each position independently and therefore fails to exploit the correlations among the errors embedded in the exact-$t$ substitution channel structure. The decoder in this section processes each column of a given output-set one at a time, in some order $\pi(1),\pi(2),...,\pi(k)$, carrying information from the previously decoded positions over to the remaining ones. The order of processing is explained in detail in Section~\ref{subsec:order}.

Suppose the first $\ell$ columns have been decided. If the decisions made were correct, then $\mathbf{y}_i$ still carries exactly $t - \sum_{m=1}^{\ell} \mathds{1}\{\mathbf{y}_i[\pi(m)] \neq \hat{\mathbf{u}}[\pi(m)]\}$ errors among the remaining undecided positions. This motivates the following definition.

\begin{definition}\label{def:budget}
Let the decoder process columns in some order $\pi(1), \pi(2), \ldots, \pi(k)$. After deciding the first $\ell$ columns, the \emph{remaining error budget} of $\mathbf{y}_i$ is
\begin{align} \label{eq:budget}
    s_i^{(\ell)} \triangleq t - \sum_{m=1}^{\ell} \mathds{1}\{\mathbf{y}_i[\pi(m)] \neq \hat{\mathbf{u}}[\pi(m)]\}
\end{align}
\end{definition}

\noindent In other words, $s_i^{(\ell)}$ is the decoder's estimate of the number of errors that $\mathbf{y}_i$ must still contain among the $r^{(\ell)}\triangleq k-\ell$ remaining undecoded columns. If all previous columns were decoded correctly, then $s_{i}^{(\ell)}$ equals the exact number of errors that $\mathbf{y}_i$ has in the remaining columns; otherwise it is only an erroneous estimate of that number. The true number of remaining errors lies in $\{0,1,\ldots,r^{(\ell)}\}$ so a \emph{valid} budget must satisfy
\begin{align}
\label{eq:budget_bounds}
    0\le s_{i}^{(\ell)}\le r^{(\ell)}.
\end{align}

An output with small remaining budget relative to~$r^{(\ell)}$ is believed to have more correct entries than errors among the remaining undecoded columns and therefore is more reliable, while one with a large budget has more errors than correct entries in those columns, and is less reliable. To capture this, we assign each output a weight 
\begin{align}
    \label{eq:weight}
    w_i^{(\ell)}\triangleq r^{(\ell)}-2s_i^{(\ell)}, 
\end{align}

Assuming correct processing history, output $\mathbf{y}_i$ has $s_i^{(\ell)}$ errors and $r^{(\ell)}-s_i^{(\ell)}$ correct entries among the~$r^{(\ell)}$ remaining columns so $w_i^{(\ell)}$ is exactly the difference between the number of correct entries and the number of errors over those columns. This weight is positive when correct entries outnumber errors, negative when errors outnumber correct entries, and zero when the two are balanced.

The next column to be decided, $j=\pi(\ell+1)$, receives the \textit{weighted vote}
\begin{align}
    \label{eq:weighted_vote}
    v^{(\ell)}_j \triangleq \sum_{i=1}^{N} w_i^{(\ell)} \cdot (-1)^{\mathbf{y}_i[j]}.
\end{align}
The decoder sets $\hat{\mathbf{u}}[j]=0$ if $v^{(\ell)}_j>0$ and $\hat{\mathbf{u}}[j]=1$ if $v^{(\ell)}_j<0$. A tie, $v^{(\ell)}_j=0$, implies that the weighted evidence is evenly split; note that this may happen even when~$N$ is odd. In such a case the decoder declares a failure, similarly to the erasure convention of the bit-wise majority-vote decoder in \eqref{eq:majrule}.

The choice to apply linear weights is arbitrary and future work can look at alternatives; however, note that by the design the contribution of each~$\mathbf{y}_i[j]$ is identical to that which would have been assigned to its complement, had the error budget been complementary to~$r^{(\ell)}$. In other words,
\begin{align*}
    &\big(r^{(\ell)}-2s_i^{(\ell)}\big) \cdot (-1)^{\mathbf{y}_i[j]}\\ 
    &= \Big(r^{(\ell)}-2\big(r^{(\ell)}-s_i^{(\ell)}\big)\Big) \cdot (-1)^{1\oplus\mathbf{y}_i[j]}.
\end{align*}

Initially, $\ell=0$ and every row has the same budget $s_i^{(0)}=t$ so the weights $w_i^{(0)}=k-2t$  are identical for all outputs. Since $t<\frac{k}{2}$, this common weight is strictly positive and the vote reduces to
\begin{align}\label{eq:first_step_vote}
    v^{(0)}_j = (k-2t)(N - 2c_j),
\end{align}
whose sign is that of $N-2c_j$. The first decision is therefore equivalent to simple bit-wise majority-vote on column $j$. The weighting takes effect from the second step onward. This observation is used in following subsections. In particular, our decoding order $\pi$ will set $\pi(1) = \operatorname{argmax}_j |v^{(0)}_j|$, a rule we will see in Section~\ref{subsec:order} generalises further.

Once all columns are decided, the decoder accepts $\hat{\mathbf{u}}$ only if $\dH(\mathbf{y}_i,\hat{\mathbf{u}})=t$ for all $i \in [N]$, and declare a failure otherwise. This condition is necessary for $\hat{\mathbf{u}}=\mathbf{u}$ under any decoder, and also sufficient whenever $N\geq \Nwc$, as discussed in Remark~\ref{rem:t-codition}.

\subsection{The First Decision}
\label{subsec:firstcol}
In subsection~\ref{subsec:budget} we showed that all output weights are identical at the first step, and therefore the decoder in this section makes the same first decision by applying the bit-wise majority-vote rule. This first decision is crucial for the process, as it cannot benefit from updated reliability belief from any previous stage. 
In this section we analyse this first decision, with the intuition that while correct decoding is maintained, subsequent decisions are biased towards being correct.

\begin{lemma}\label{lem:first-position}
If $N > \frac{\binom{k-1}{t-1}}{1-\frac{t}{k}}$ then position $\pi(1)$ is decoded correctly by majority-vote.
\end{lemma}
\begin{IEEEproof}
Observe that for all $j\in[k]$ we have $c_j\leq \binom{k-1}{t-1}$, as at most that number of distinct outputs can be erroneous at position~$j$. This implies that the maximum reliability of a position with weight $w^{(0)}_j<0$ is $|v^{(0)}_j| \sim |N-2\binom{k-1}{t-1}|$. ($\sim$ is used here to denote ignoring the multiplicative factor $(k-2t)$ shared by all columns in the first iteration.)

Conversely, since every output contains $k-t$ error-free positions, counting across all outputs yields $N(k-t)$ such positions overall; by the pigeon-hole principle, for at least a single $j\in[k]$ it holds that $c_j\leq N-\frac{N(k-t)}{k}$. 
Thus, when 
\begin{align}\label{eq:pigeon-new}
    \frac{N}{2} - \left(N-\frac{N(k-t)}{k}\right) > \binom{k-1}{t-1} - \frac{N}{2}, 
\end{align}
we have $|v^{(0)}_j| > \max\lbrace|v^{(0)}_{j'}| : w^{(0)}_{j'}<0\rbrace$, thus the most reliable position, the first to be decoded, is decoded correctly. 
Rearranging \eqref{eq:pigeon-new} gives the claim.
\end{IEEEproof}

\begin{corollary}
For $t=2<\frac{k}{2}$ and $N=\Nwc=2k-3$, the column with highest reliability is decoded correctly by majority-vote.
\end{corollary}
\begin{IEEEproof}
Note that $\Nwc > \frac{\binom{k-1}{t-1}}{1-\tfrac{t}{k}}$ if and only if $(2k-3)(1-\frac{2}{k}) > k-1$, and by reordering 
\begin{align*}
    k^2-6k+6 > 0.
\end{align*}
This holds for all $k\geq 5$.
\end{IEEEproof}

Motivated by the above lemma, we make the following notation:
\begin{align}\label{eq:Nwmaj}
    \Nwmaj \triangleq \bigg\lfloor\frac{\binom{k-1}{t-1}}{1-\tfrac{t}{k}}\bigg\rfloor+1.
\end{align}
The question of whether the inequality of Lemma~\ref{lem:first-position} is sufficient for the designed algorithm to successfully decode \emph{all} positions is left unanswered. However, we present empirical results in Table~\ref{tab:backtracking_performance} supporting the affirmative. It also contains limited evidence that the inequality is tight, i.e., at $N=\Nwmaj-1$ there exist output-sets for which the proposed algorithm fails to reconstruct the transmitted sequence.

\subsection{Decoding order}\label{subsec:order}

\begin{algorithm}[t]
\caption{Sequential Weighted Majority-Vote Decoder}\label{alg:weighted}
\begin{algorithmic}[1]
\REQUIRE Sequence matrix $\mathbf{Y} \in \Ftwo^{N \times k}$, error count $t$
\ENSURE Estimated sequence $\hat{\mathbf{u}} \in \Ftwo^k$ or FAILURE
\STATE $s_i \leftarrow t$ for all $i \in [N]$
\STATE $r \leftarrow k$
\FOR{$\ell = 0, 1, \ldots, k-1$}
    \STATE $j \leftarrow \pi(\ell+1) = \operatorname*{argmax}_{j\notin \pi([1,\ell])} \big|v^{(\ell)}_j\big|$ w. tie-breaking
    \STATE $w_i \leftarrow r - 2s_i$ for all $i \in [N]$
    \STATE $v_j \leftarrow \sum_{i=1}^{N} w_i \cdot (-1)^{\mathbf{y}_i[j]}$
    \IF{$v_j > 0$}
        \STATE $\hat{\mathbf{u}}[j] \leftarrow 0$
    \ELSIF{$v_j < 0$}
        \STATE $\hat{\mathbf{u}}[j] \leftarrow 1$
    \ELSE
        \IF{can backtrack}
            \STATE Backtrack 
        \ELSE
            \RETURN FAILURE 
        \ENDIF
    \ENDIF
    \STATE $s_i \leftarrow s_i - \mathds{1}\{\mathbf{y}_i[j] \neq \hat{\mathbf{u}}[j]\}$ for all $i \in [N]$
    \STATE $r \leftarrow r - 1$
    \IF{$s_i < 0$ or $s_i > r$ for some $i$}
        \IF{can backtrack}
            \STATE Backtrack 
        \ELSE
            \RETURN FAILURE 
        \ENDIF
    \ENDIF
\ENDFOR
\IF{$\dH(\mathbf{y}_i, \hat{\mathbf{u}}) = t$ for all $i \in [N]$}
    \RETURN $\hat{\mathbf{u}}$
\ELSE
    \IF{can backtrack}
        \STATE Backtrack 
    \ELSE
        \RETURN FAILURE 
    \ENDIF
\ENDIF
\end{algorithmic}
\end{algorithm}

As mentioned earlier, reliability is the measure we use to determine decoding order. We quantify it by the absolute value of each as-of-yet-undecoded position's vote, $|v^{(\ell)}_j|$ at each step $(\ell+1)$. That is, 
\begin{align*}
    \pi(\ell+1)\triangleq \operatorname*{argmax}_{j\notin \pi([1,\ell])} \big|v^{(\ell)}_j\big|, 
\end{align*}
where $\pi([1,\ell])\triangleq \lbrace \pi(i) : 1\leq i\leq \ell\rbrace$, and is empty for $\ell=0$ (agreeing with our earlier definition of $\pi(1)$). 
Ordering the columns by decreasing reliability lets the decoder make its most confident decision first, obtaining accurate budget updates that benefit the decisions for less reliable columns that follow.

When all column reliabilities are distinct, the permutation $\pi(\cdot)$ is uniquely determined by the input. However, when at least two columns share a reliability value at any step, a case we name \textit{reliability degeneracy}, their order is ambiguous. Different orderings can influence the decoder's success-rate. The following example illustrates the impact of this tie-braking convention. 

\begin{example}
\label{ex:degeneracy}
Consider $k=5$, $t=2$, $N=\Nwc=7$, with $\mathbf{u}=\mathbf{0}$, and the output sequence matrix 
\begin{equation*}
    Y\triangleq
    \begin{pmatrix}
         1&1&0&0&0\\1&0&1&0&0\\1&0&0&1&0\\1&0&0&0&1\\0&1&1&0&0\\0&1&0&1&0\\0&0&1&1&0
    \end{pmatrix} \in Y_7(5,2)
\end{equation*}
The column sums are $(4,3,3,3,1)$ and the corresponding reliabilities $\sim(1,1,1,1,5)$. Column~$5$ is decoded first (highest reliability) and correctly yields $\hat{\mathbf{u}}[5]=0$. After this step, row~$4$ has budget $s_4^{(1)}=1$ while all other rows have budget $s_i^{(1)}=2$. The remaining four columns form a reliability-degenerate group. At this point, only row~$4$ has nonzero weight ($w_4^{(1)}=4-2=2$, while $w_i^{(1)}=4-4=0$, for $i\neq4$). If column~$1$ is decoded next (as per the tie-breaking column index convention), row~$4$'s entry $\mathbf{y}_4[1]=1$ dominates the weighted vote, yielding $v^{(1)}_1=-2<0$ and the incorrect decision $\hat{\mathbf{u}}[1]=1$. However, if any of the columns $2$,$3$, or $4$ is decoded next, row $4$'s entry $\mathbf{y}_4[j]=0$ yields $v^{1}_j=2>0$ and the correct decision $\hat{\mathbf{u}}[j]=0$.
\end{example}

Motivated by this fact, we insert a backtracking rule, that allows the decoder to reprocess cases of reliability degeneracy. These cases are handled by a tie-breaking convention following column order, but whenever the decoder reaches a failure (by either violation of \eqref{eq:budget_bounds}, termination violating the condition of Remark~\ref{rem:t-codition}, or draw vote $v^{(\ell)}_{\pi(\ell+1)}=0$) it backtracks to the last-made tie-breaking and traverses the next available choice. Failure is now declared only when no unspent ties remain. 

The resulting decoder is given in Algorithm~\ref{alg:weighted}. Table~\ref{tab:backtracking_performance} reports its performance, as well as that of classical bit-wise majority decoding, on all admissible inputs at $N =\{\Nwc,\ldots,\Nmaj\}$ and $k \in \{5, 6, 7, 8,9\}$, $t \in\{2,3\}$, whenever exhaustive enumeration is feasible. These empirical results give rise to the (admittedly  optimistic) Conjecture~\ref{conj:backtracking}; a rigorous proof or obtaining a counter-example are left for future work.

\begin{conjecture}\label{conj:backtracking}
\begin{enumerate}
    \item Algorithm~\ref{alg:weighted} uniquely decodes the transmitted sequence whenever $N\geq \max\lbrace\Nwc,\Nwmaj\rbrace$.
    \item Algorithm~\ref{alg:weighted} uniquely decodes the transmitted sequence whenever bit-wise majority does.
\end{enumerate}
\end{conjecture}

Whether any efficient decoder guarantees decoding given $\Nwc$ distinct outputs remains open; the results of this section instead delimit where the weighted approach reaches the threshold and where it demonstrably does not.

\begin{table*}[t]
  \centering
  \caption{Performance of weighted majority-vote decoding by exhaustive search on all admissible output-sets. In every instance of failure, bit-wise majority also failed. For each row, the total count of distinct output-sets equals $\binom{\binom{k}{t}}{N}$.}
 \label{tab:backtracking_performance}
  \small
  \setlength{\tabcolsep}{5pt}
  \begin{tabular}{c c l r r r}
    \toprule
    $k$ & $t$ & $N$ & \makecell{Total no.\\of  output-sets} 
        & \makecell{Alg.~\ref{alg:weighted}\\ failures} & \makecell{Bit-wise majority\\failures} \\
        
    \midrule
    \multirow{2}{*}{5} & \multirow{2}{*}{2} & 7$\>=\Nwc=\Nwmaj$ & 120 & 0 & 90 (75\%) \\
     &  & 9$\>=\Nmaj$ & 10 & 0 & 0 \\
    
    \midrule
    
    \multirow{2}{*}{6} & \multirow{2}{*}{2} & 9$\>=\Nwc>\Nwmaj$ & 5{,}005 & 0 & 1{,}245 (24.88\%) \\
     &  & 11$\>=\Nmaj$ & 1{,}365 & 0 & 0 \\
    
    \midrule
    
    \multirow{8}{*}{7} & \multirow{2}{*}{2} & 11$\>=\Nwc>\Nwmaj$ & 352{,}716 & 0 &  21{,}000 (5.95\%) \\
     &  & 13$\>=\Nmaj$ & 203{,}490 & 0 & 0 \\
    
    \arrayrulecolor{black!35}    
    \cmidrule{2-6}
    \arrayrulecolor{black} 
    
     & \multirow{6}{*}{3} & 21$\>=\Nwc$ & 2{,}319{,}959{,}400 & 560{,}490 (0.0242\%) &  1{,}819{,}700{,}190 (78.44\%) \\
     &  & 23 & 834{,}451{,}800 & 13{,}440 (0.00161\%) & 566{,}796{,}300 (67.92\%) \\
     &  & 25 & 183{,}579{,}396 & 42 (0.0000229\%) & 98{,}888{,}146 (53.87\%) \\
     &  & 27$\>=\Nwmaj$ & 23{,}535{,}820 & 0 & 8{,}516{,}445 (36.19\%) \\
     &  & 29 & 1{,}623{,}160 & 0 & 266{,}910 (16.44\%) \\
     &  & 31$\>=\Nmaj$ &  52{,}360 & 0 & 0 \\
    
    \midrule
    
    \multirow{4}{*}{8} & \multirow{2}{*}{2} & 13$\>=\Nwc>\Nwmaj$ & 37{,}442{,}160 & 0 & 434{,}084 (1.16\%) \\
     &  & 15$\>=\Nmaj$ & 37{,}442{,}160 & 0 & 0 \\
    
    \arrayrulecolor{black!35}   
    \cmidrule{2-6}
    \arrayrulecolor{black}

     & \multirow{2}{*}{3} & 41$\>>\Nwc=31$ & 16{,}253{,}249{,}498{,}640 & --- & --- \\
     &  & 43$\>=\Nmaj$ & 1{,}889{,}912{,}732{,}400 & 0 & 0 \\
    
    \midrule
    
    \multirow{2}{*}{9} & \multirow{2}{*}{2} & 15$\>=\Nwc>\Nwmaj$ & 5{,}567{,}902{,}560 & 0 & 10{,}656{,}324 (0.19\%) \\
     &  & 17$\>=\Nmaj$ & 8{,}597{,}496{,}600 & 0 & 0 \\

    \bottomrule
  \end{tabular}
\end{table*}

%% file: chapters/7_conc.tex
\label{sec:conc}
In this paper we revisit the exact-$t$ substitution error channel model and characterize its worst-case guarantee of unique reconstruction, for binary sequences. 
Using this result, we demonstrate that the bit-wise majority-vote decoder, proven by Levenshtein to be optimal in the up-to-$t$ scenario, is in fact suboptimal here. 
However, due to its simplicity, we devise an early-termination rule for the bit-wise majority-vote that applies even in cases where less than the worst-case required number of distinct channel output observations are available to the decoder. 

We then devise a weighted-majority decoder that exploits the correlations among errors in an attempt to improve the threshold of observations for unique reconstruction. This decoder comes with modest complexity increase, and outperforms simple bit-wise majority-vote decoding, numerically. 
We demonstrate the resulting performance increase of that decoder in a number of cases, which supports our (optimistic) conjecture that the suggested decoder indeed lowers the number of required observations to guarantee worst-case reconstruction.